\documentclass[sn-aps]{sn-jnl} 

\usepackage{graphicx}
\usepackage{bm,amsmath,amssymb}
\usepackage{xcolor}
\usepackage{longtable,booktabs}
\usepackage{siunitx}
\usepackage{sidecap}

\DeclareMathOperator{\psin}{polsin}

\newcommand{\abs}[1]{\ensuremath{\mid#1\mid}}

\newcommand{\void}[1]{}

\jyear{2021}

\begin{document}

\title{Machine-learning correction to density-functional crystal structure optimization}

\author*[1,2]{\fnm{Robert} \sur{Hussein}}    
\author[3,2]{\fnm{Jonathan} \sur{Schmidt}}
\author[3,2]{\fnm{Tom\'as} \sur{Barros}}
\author[3,2]{\fnm{Miguel A.L.} \sur{Marques}}
\author[1,2]{\fnm{Silvana} \sur{Botti}}

\affil[1]{Institut f\"ur Festk\"orpertheorie und -optik, Friedrich-Schiller-Universit\"at Jena, D-07743 Jena, Germany}
\affil[2]{European Theoretical Spectroscopy Facility}
\affil[3]{Institut f\"ur Physik, Martin-Luther-Universit\"at Halle-Wittenberg, D-06099 Halle, Germany}

\abstract{
Density functional theory is routinely applied to predict crystal structures. The most common exchange-correlation functionals used to this end are the Perdew-Burke-Ernzerhof (PBE) approximation and its variant PBEsol. We investigate the performance of these functionals for the prediction of lattice parameters and show how to enhance their accuracy using machine learning. Our dataset is constituted by experimental crystal structures of the Inorganic Crystal Structure Database matched with PBE-optmized structures stored in the materials project database. We complement these data with PBEsol calculations. We demonstrate that the accuracy and precision of PBE/PBEsol volume predictions can be noticeably improved \textit{a posteriori} by employing simple, explainable machine learning models. These models can improve PBE unit cell volumes to match the accuracy of PBEsol calculations, and reduce the error of the latter with respect to experiment by \SI{35}{\percent}. Further, the error of PBE lattice constants is reduced by a factor of 3--5. A further benefit of our approach is the implicit correction of finite temperature effects without performing phonon calculations. 
}


\maketitle

\section{Introduction}
Computational high-throughput studies form the basis for the discovery of new materials in  modern material science. In solid state physics, these studies are mostly 
performed within Kohn-Sham density functional theory (DFT)~\cite{HohenbergPR1964a,KohnPR1965a,JonesRMP1989a}. While DFT  formally  provides an exact description 
of the many-body Schr\"odinger equation, it relies in practice on approximations for the exchange-correlation energy. In solid-state physics, one commonly utilizes the
Perdew-Burke-Ernzerhof's PBE functional~\cite{PerdewPRL1996a}. While PBE and its variants are successful in predicting structural and electronic properties of solids,  
they may yield, nevertheless, non-negligible deviations from experiments. Specifically, PBE underestimates atomic bond lengths, thus, overestimating lattice 
constants~\cite{TranPRB2007a,ZhangNewJP2018a,KovacsJCP2019a} and volumes. Variants of PBE such as the PBE for solids (PBEsol) ~\cite{PerdewPRL2008a} 
were designed to improve upon this problem~\cite{HaasPRB2009a}. However, they still do not achieve the desired accuracy in comparison with experiments.

The correctness of the lattice constants 
and the corresponding unit cell volumes is indispensable for a reliable prediction of bulk electronic properties ~\cite{MurnaghanPNAS1944a,BirchPR1947a,ZiambarasPRB2003a} and 
when considering experimental realizations of composite materials. For instance, the lattice mismatch between growth substrates and films can be a source of major problems in experiments.
Another reason to focus on lattice parameters is the fact that this is the material property for which it is possible to find the largest amount of experimental data, collected in the Inorganic Crystal Structure Database (ICSD)~\cite{Bergerhoff1987a}.

In this work, we demonstrate that machine learning methods can improve the lattice volume predictions based on PBE/PBEsol without increasing the computational effort. 
Machine learning enjoyed over the past few years great success in a wide variety of applications~\cite{SchmidtnpjCM2019a} ranging from property predictions of 
bandgaps~\cite{GuSolidStateS2006a, PilaniaSR2016a, ZhuoJPCL2018a} and elastic moduli~\cite{JongSR2016a} to the stability analysis of crystals~\cite{YeNC2018a,SchmidtJCP2018a} 
and molecular force field estimations~\cite{GlielmoPRB2018a}. Recently, the prediction of lattice constants and volumes generated much 
interest~\cite{JavedCMS2007a, TakahashiJCP2017a, MajidCMS2010a, ZhangCPL2020a, NaitAmarJPCB2020a, AladeJAP2020a, LiArXiv2020a, ZhangChemistrySelect2020a, ZhangIJACT2021a}. 
The majority of these studies is, however, limited to a particular crystal structure. In contrast to previous studies which are mainly based on direct calculations, we are building our approach on first-principles calculations, aiming at improving their accuracy in comparison with corresponding experiments. Our approach is not  limited to a specific crystal structure or a subset of chemical elements. We will focus here on applying explainable machine learning methods~\cite{LiptonQueue2018a,DuCACM2019a} to correct errors of PBE/PBEsol calculations of crystal structures of newly predicted materials. 
Specifically, we will employ model agnostic supervised local explanations (MAPLEs)~\cite{Plumb2018a} in combination with a random forest model~\cite{BreimanMachineLearning2001a} 
to combine the high accuracy of tree models for small datasets and the interpretability of MAPLE models.
\begin{SCfigure}
\includegraphics[width=0.66\textwidth]{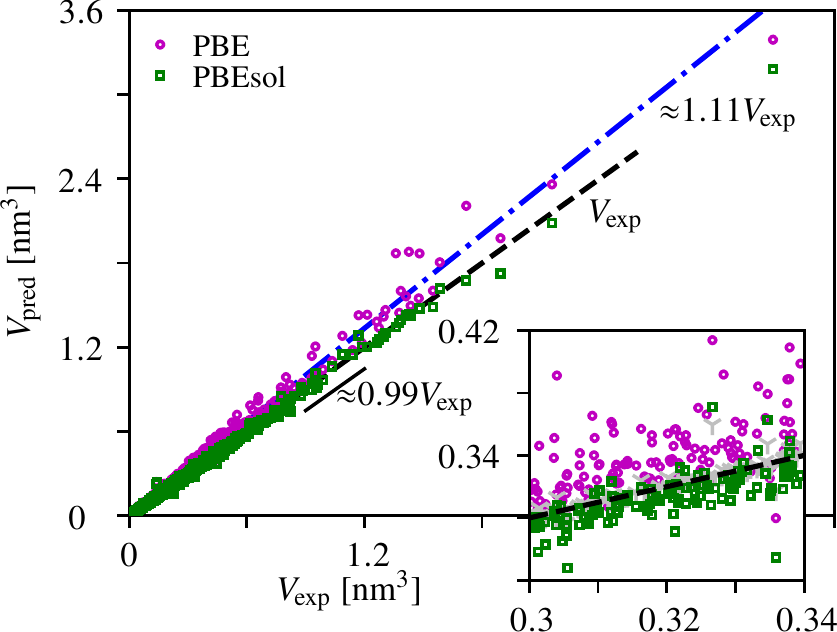}\caption{\label{fig.:overview}
(a) Correlation between measured  primitive unit cell volumes $V_{\rm exp}$ and predicted ones $V_{\rm pred}$. Ideal predictions match the black dashed line. 
The blue dash-dotted line depicts PBE's linear regression. The inset additionally considers the machine learning prediction ${\rm pred= PBE{+}MAPLE}$ marked by gray tripods. 
}
\end{SCfigure}

Before diving into detail, we can observe in figure~\ref{fig.:overview} the primitive unit cell volumes  $V_{\rm pred}$, predicted from DFT calculations using PBE and PBEsol functionals, plotted against the experimental unit cell volumes $V_{\rm exp}$. We remark that the primitive cell volume is the simplest quantity that can be directly compared, independently of the specific details of the crystal structure and chemical composition. 
Their correlation gives a first impression about the accuracy (systematic error) and precision (variability) of 
the theoretically estimated unit cell volumes. Calculations with the PBE functional (magenta circles) significantly overestimates the measured volumes by roughly \SI{11}{\percent}, while PBEsol (green squares) provides a much better approximation of them.

On a closer look, one sees that $V_{\rm exp}$ constitutes a soft lower bound on the PBE volumes in the sense that about \SI{90}{\percent} of the predicted volumes lay above it. This is a consequence of the tendency of PBE to underbind. This soft bound entails a skewness on the predicted PBE volume distribution, which we revisit later. The inset of figure~\ref{fig.:overview} shows a close-up view of primitive unit cell volumes of about \SI{0.32}{\cubic\nano\meter} to better distinguish the individual data points. We additionally include in the inset  the volumes obtained by correcting PBE calculations with machine learning (gray tripods) to anticipate visually the strong error reduction. We will discuss thoroughly the machine-learning corrections in the next sections.

The remaining article is organized as follows. In section~\ref{sec.:model}, we present the employed machine learning models. Details on the experimental and theoretical datasets
and their matching are discussed in Sec.~\ref{sec.:data}. We analyze in section~\ref{sec.:analysis} our predictive models and compare their performance with the one of underlying DFT calculations. 
In section~\ref{sec.:lattice}, we discuss the correction of the lattice constants. The last section is devoted to our conclusions.

\section{\label{sec.:model}Predictive models}
Tree-ensemble-based models such as random forests~\cite{BreimanMachineLearning2001a} and gradient boosting \cite{FriedmanSM2003a,Theodoridis2015a}  
are known to be suitable to the description of material properties for relatively small datasets~\cite{IsayevCM2015a,FurmanchukRSCA2016a} but they are not restricted to them~\cite{StanevnpjCM2018a}.
A drawback of employing multiple-decision trees is however their general lack of interpretability~\cite{DuCACM2019a}. Appropriate  combination with local linear 
models~\cite{ClevelandJASA1988a,RuppertAS1994a,Barrientos-MarinJNS2010a}, as in model agnostic supervised local explanations (MAPLEs)~\cite{Plumb2018a},  
overcomes this deficiency by  providing local and example-based explanations. The former addresses causal relations between specific \textit{input features} of an individual 
prediction (such as lattice constants) and its outcome by identifying their importances~\cite{BaehrensJMLR2010a,Plumb2018a,LundbergNMI2020a}. The latter asks instead 
for the  contribution of specific \textit{training points}~\cite{CookTechnometrics1980a,Kim2016a,BienAAS2011a}.

In this work, we employ the MAPLE implementation of Plumb et al.~\cite{Plumb2018a} as well as tree models and utility functions from Ref.~\cite{PedregosaJMLR2011a}. 
We evaluate the machine learning models by tenfold Monte-Carlo cross-validation. We choose this approach instead of using an independent testset,  since our full dataset 
exhibits a large variance in structures/elements while being relatively small in size. In each independent run of the cross-validation scheme, the full dataset is randomly split at a ratio $1:9$
without replacement into a test and a training set. The hyper-parameter optimization has been performed on a separate random splitting.
Here, the number of trees forming the tree ensemble turns out to be the most important hyper parameter. The minimal number of samples controlling the splitting is of minor importance.
The theoretical crystal structures calculated using DFT at the PBE and PBEsol level, respectively, serve as input parameters for the training, complemented with composition-specific features provided by Matminer~\cite{WardCMS2018a}.

By training the MAPLE models with the datasets discussed in Sec.~\ref{sec.:data}, we find that the primitive-cell volume prediction is, indeed, to a large extent based on structural quantities, see Table~\ref{tab.:feature} in the Appendix.
Here, we use the average of the root-node impurity over the decision-tree ensemble as an estimator to quantify the relevance of the features.
The binary splittings of an individual  decision tree are such constructed that they minimize its impurity. In this sense, the first splitting and,
therewith, the respective root-node feature has a major impact on the decision-tree structure and the model prediction.
In particular, the splitting of the training set with respect to the root feature is in more than \SI{50}{\percent} of the cases directly related to the crystal structure of the compounds, through, e.g., the volume $V$, the 
lattice constants  $a$, $b$, $c$ and the corresponding angles $\alpha$, $\beta$, $\gamma$.
Concerning the compositional features, the periodic table based features and averaged thermal properties of the elements play by far the largest role exceeding also two of the lattice angles in importance.
Our models are available at Ref.~\cite{code}.

\section{\label{sec.:data} Dataset}
For our analysis and model training we consider roughly $2000$ PBE-structures from the Materials Project (MP)~\cite{JainAPLM2013a}. The corresponding PBEsol calculations 
are available from Ref.~\cite{SchmidtMaterialsCloudArchive2021a}.
The experimental crystal structures are extracted from the ICSD~\cite{Bergerhoff1987a}. 
A mapping of ICSD- and MP-identifiers is provided in Table~\ref{tab.:mapping} of the Appendix. 

We remark that experiments are conducted at finite temperature (\SIrange{2}{373}{\kelvin}) and pressure ($\leq$\SI{1}{\bar}) while DFT calculations describe equilibrium structures at zero temperature and pressure. In order to obtain PBE(sol) crystal structures in the same thermodynamic conditions than experimental samples they should be corrected by expensive phonon calculations for  thermal expansions and zero-point effects arising from finite lattice fluctuations~\cite{GrabowskiPRB2007a,LejaeghereCRSolidStateMS2014a,AllenPRB2015a,RitzJAP2019a}. 
For small molecules, the ambient pressure has additionally to be taken into account~\cite{HojaWIREsCMS2017a} but may be neglected for solids~\cite{GrabowskiPRB2007a}.
By training the predictive models on finite temperature volumes $V_{\rm exp}$ as target variables
one has the advantage to implicitly include finite temperature corrections. In principle, the ambient temperature of the measurements could be included 
as an input parameter for machine learning. However it turns out that the resulting models are mostly independent of temperature, since the large majority of the experiments are performed 
at roughly the same temperature (about~\SI{293}{\kelvin}, median and mode). The mean value of the temperature distribution is indeed \SI{271}{\kelvin}.
\begin{SCfigure}
\includegraphics[width=0.66\textwidth]{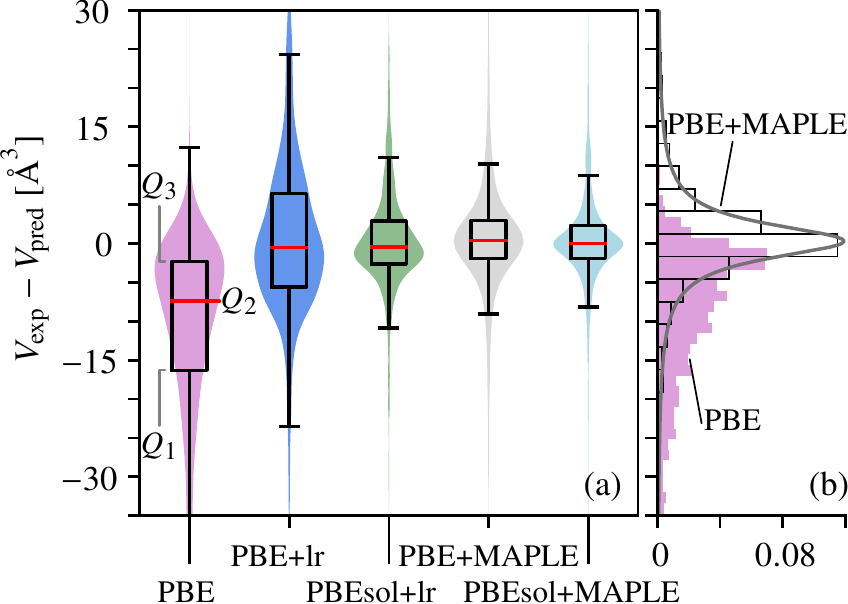}\caption{\label{fig.:residuals}
Violin plot (a) and probability densities (b) of the volume residuals $V_{\rm exp}-V_{\rm pred}$ for the indicated testsets. The suffix `+lr' indicates linear regression and $Q_k$ is the $k$-th quartile.
}
\end{SCfigure}

\section{\label{sec.:analysis}Volume predictions}
In this section, we compare volume predictions obtained with various machine learning models. First, we give an overview of these predictions 
by discussing the central characteristics of their volume residuals. Then, we study their cross-validation error and address finally the convergence of the model training.

We show in figure~\ref{fig.:residuals}(a) violin plots~\cite{HintzeAS1998a} of the volume residuals $V_{\rm exp}-V_{\rm pred}$. 
One sees clearly that the median of the DFT-PBE calculations (red line in magenta violin) of about $Q_2\approx\SI{-7.4}{\cubic\angstrom}$ is, 
indeed, corrected by simple linear regression (blue violin). Also its interquartile range ${\rm IQR} \equiv Q_3-Q_1$ of about \SI{14}{\cubic\angstrom} is roughly reduced by \SI{2}{\cubic\angstrom}  
with the drawback that already well predicted volumes worsen. The skewness remains. Employing MAPLE cures the skewness and reduces 
the spreading further up to a third of the initial value (see gray violin). Intriguingly, its volume forecast is comparable to the simple linear regression prediction starting from PBEsol volumes.
Beyond the median and the interquartile range, the violin shapes in Fig.~\ref{fig.:residuals}(a) estimate the entire probability densities 
of the volume residuals. For the purpose of assessing their estimation quality, we compare the DFT-PBE estimate with the corresponding normalized histogram depicted in 
Fig.~\ref{fig.:residuals}(b). The estimate captures well the curve progression but is less pronounced around its mode located at \SI{-1.3}{\cubic\angstrom}. 
Additionally, we show in panel (b)  the normalized histogram of the MAPLE prediction that corrects PBE volumes. As prefigured, it is considerably narrower and can be well approximated by a slightly biased Lorentzian (gray solid line) with a linewidth  of roughly its  interquartile range. 
\begin{SCfigure}
\includegraphics[width=0.66\textwidth]{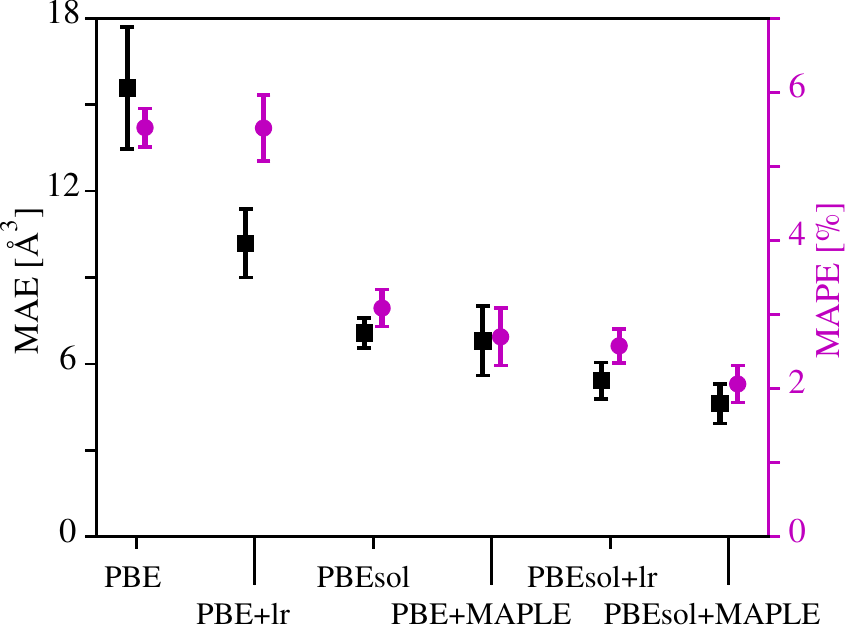}\caption{\label{fig.:benchmarkVolumes} Cross-validation errors for different models with standard deviations as error bars. 
}
\end{SCfigure}

For a more quantitative comparison of the different models, we focus in the following on the cross-validation error.
The cross-validation error of a specific model, is obtained by evaluating for each testset the error of its prediction with respect to the measured value, and taking the arithmetic mean of these errors. Additionally, we determine the standard deviation of the individual errors. As error metrics we chose the mean absolute error $\mathrm{MAE}=\sum_{k=1}^n \abs{y_{k,\rm exp}-y_k}/n$ and the mean absolute percentage error 
$\mathrm{MAPE} =100\sum_{k=1}^n\abs{y_{k,\rm exp}-y_k}/\abs{n y_{k,\rm exp}}$, where $y_{k,\rm exp}$ ($y_k$) indicates the measured (predicted) property of 
the $k$-th sample.  Since all experimental volumes are finite, $\mathrm{MAPE}$ is well defined.
In figure~\ref{fig.:benchmarkVolumes}, we show the cross-validation errors of the predicted primitive unit-cell volumes for different models. 
Their numerical mean values and standard deviations are listed in Table~\ref{tab.:crossValErr}.
As expected, DFT-PBE itself leads overall to the worst cross-validation errors while PBE corrected with linear regression (+lr) improves the MAE leaving the MAPE unchanged. The MAPLE 
model based on PBE volumes reduces the PBE-MAE by about~\SI{50}{\percent} and is slightly better than DFT-PBEsol volumes. However, PBEsol corrected with linear regression is 
once again better. Most importantly, the MAPLE model based on PBEsol shows the best MAE improving by roughly~\SI{35}{\percent} upon DFT calculations alone with this functional. 
The IQRs in Table~\ref{tab.:crossValErr} show the same tendency as the MAE and MAPE, supporting the conclusion regarding the possible improvements achievable with the MAPLE models.
Additionally, we report therein the MAPLE models using the measured temperature as an input feature. They perform, however, very similarly to the 
models that do not include such feature, as expected in view of the fact that most experiments were conducted at about the same temperature.
\begin{table}[t]
\begin{center}
\begin{minipage}{0.74\textwidth}
\caption{\label{tab.:crossValErr}Cross-validation errors of volume predictions and standard deviations for different DFT functionals and correction models.
The superscript `$\ast$' indicates the corresponding models including the   temperature of the measurement as an additional feature.}
\begin{tabular}{@{}rrrrrrr@{}}
\toprule
& \multicolumn{2}{c}{MAE [\si{\cubic\angstrom}]} & 
   \multicolumn{2}{c}{MAPE [\si{\percent}]} &
   \multicolumn{2}{c}{IQR [\si{\cubic\angstrom}]}\\
\cmidrule(rl){2-3} \cmidrule(rl){4-5}\cmidrule(rl){6-7}
    model & mean & std & mean & std & mean & std\\
\midrule
PBE & $15.6$ & $2.1$ & $5.5$ & $0.3$ & $14.0$ & $0.7$\\
PBE+lr & $10.2$ & $1.2$ & $5.5$ & $0.4$ & $11.9$ & $1.1$\\
PBEsol & $7.1$ & $0.5$ & $3.1$ & $0.3$ & $6.9$ & $0.7$\\
PBE+MAPLE & $6.8$ & $1.2$ & $2.7$ & $0.4$ & $5.1$ & $0.8$\\
$\text{PBE+MAPLE}^\ast$ & $6.6$ & $0.9$ & $2.6$ & $0.3$ & $4.9$ & $0.4$\\
PBEsol+lr & $5.4$ & $0.6$ & $2.6$ & $0.2$ & $5.5$ & $0.5$\\
$\text{PBEsol+MAPLE}^\ast$ & $4.8$ & $0.7$ & $2.1$ & $0.3$ & $4.2$ & $0.6$\\
PBEsol+MAPLE & $4.6$ & $0.7$ & $2.1$ & $0.2$ & $4.2$ & $0.5$\\
\bottomrule
\end{tabular}
\end{minipage}
\end{center}
\end{table}  

\begin{SCfigure}[\sidecaptionrelwidth][b]
\includegraphics[width=0.66\textwidth]{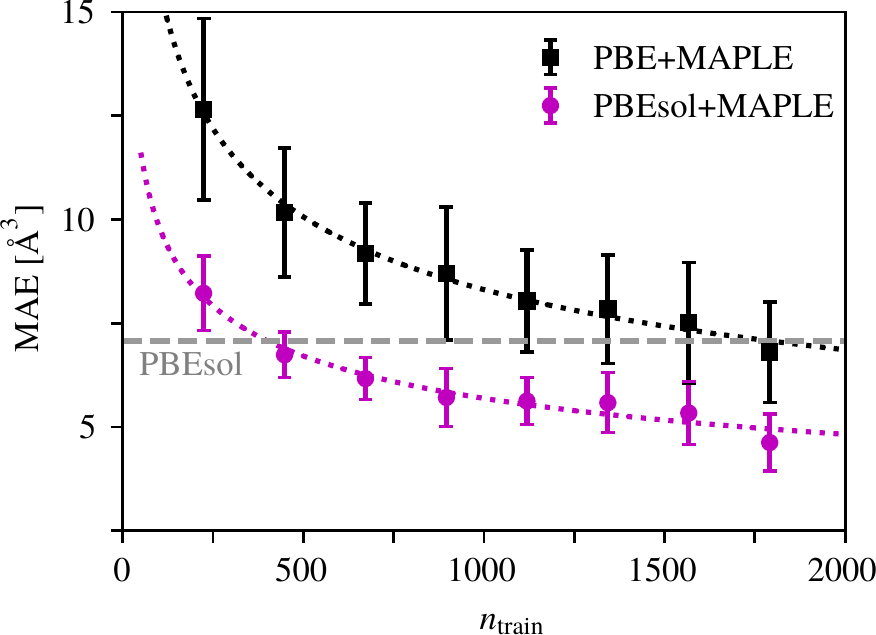}\caption{\label{fig.:conv}
Cross-validation errors for different models as a function of the training-set size $n_{\rm train}$. The error bars indicate
their standard deviation and the dotted lines correspond to their polynomial fit discussed in the main text. As reference, we indicate PBEsol's
MAE for the testset with the horizontal dashed line.}
\end{SCfigure}

To assess the learning progress of the MAPLE models, we study in figure~\ref{fig.:conv} the dependence of their MAE on the trainingset size $n_{\rm train}$.
The MAE's are again obtained by tenfold cross-validation. As expected, they decrease polynomially with the trainingset size $n_{\rm train}$~\cite{SchmidtCM2017a,ZhangnpjCM2018a}. 
In particular, we find with $\text{MAE} \propto n_{\rm train}^{-0.28}$ for PBE+MAPLE  and $\text{MAE} \propto n_{\rm train}^{-0.24}$ for PBEsol+MAPLE
a similar learning behaviour for both predictive models. Including the total data available in the ICSD, the cross-validation error could be potentially reduced by \SI{50}{\percent}.
The relatively fast decay of the cross-validation errors with respect to the training-set size makes these correction procedures already applicable for small training sets. We would 
expect it to generalize to other material properties such as bulk moduli or formation energies for which very few experimental data are available and DFT results are a worse starting point.

\begin{SCfigure}
\includegraphics[width=0.66\textwidth]{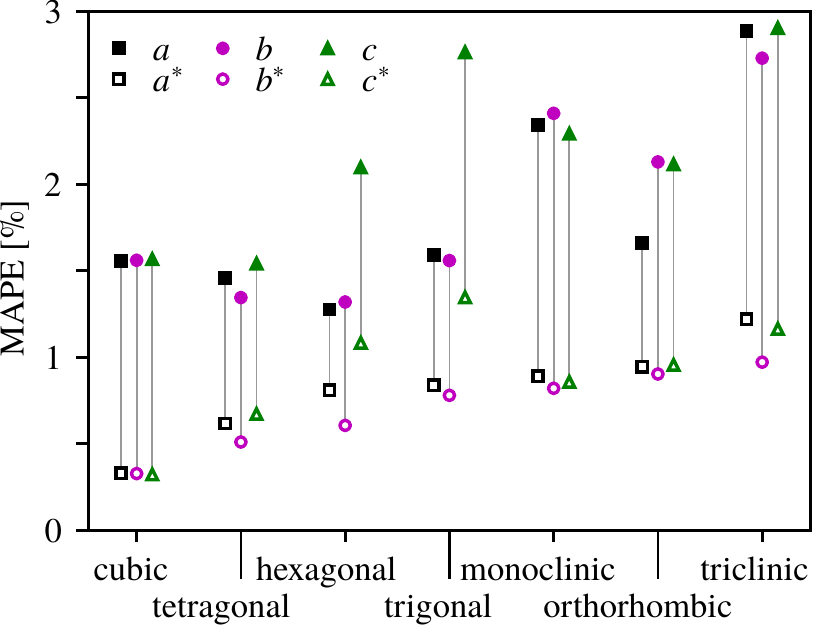}\caption{\label{fig.:lattice}
Errors of  the DFT-PBE lattice constants $a$, $b$, $c$ and the rescaled ones  $a^*$, $b^*$, $c^*$ according to Eq.~\eqref{eq.:latticeScaling} with respect to experimental values 
from the ICSD for different crystal structures. The gray vertical lines serve as a guidance for the eye for tracking the individual improvements.
}
\end{SCfigure}

\section{\label{sec.:lattice}Lattice constants} 
This far, we have discussed volume corrections. To a certain extent, we can, therewith, also improve the lattice constants as we show in this section. 
To this end, we recall how the volume is calculated.
The unit cell volume $V(\bm{a},\bm{b},\bm{c})$ is obtained from the triple product of the three lattice vectors $\bm{a}$, $\bm{b}$, $\bm{c}$ and can be written as product 
$V=abc \abs{\psin(\alpha,\beta,\gamma)}$ of a factor $abc$ only depending on their lengths and a dimensionless factor $\abs{\psin(\alpha,\beta,\gamma)}$ only depending 
on their interior angles.\footnote{The polar sine satisfies $\psin(\alpha,\beta,\gamma)\equiv\det([\bm{a},\bm{b},\bm{c}])/abc=\big(1+ 2\cos\alpha\cos\beta\cos\gamma-\cos^2\alpha-\cos^2\beta-\cos^2\gamma\big)^{1/2}$~\cite{ErikssonGeometriaeDedicata1978a}.}
The latter is for cubic crystal systems well predicted by PBE and PBEsol with a MAPE of the order of \SI{0.02}{\percent}  while lower symmetric systems does not exceed a MAPE of \SI{0.6}{\percent}. Exploiting the simplification that all lattice constants coincide for cubic crystals, 
we can extract the lattice constant correction by the prescription 
\begin{align}
	a\to a^*\equiv a \sqrt[3]{\frac{V_{\rm PBE(sol)+MAPLE}}{V_{\rm PBE(sol)}}} \label{eq.:latticeScaling}
\end{align}
for DFT calculations using  PBE(sol).  Therewith, the MAPE of the lattice vectors is roughly reduced by a factor of $5$, see figure~\ref{fig.:lattice}.
If we use the same prescription as an approximate way to correct the lattice constants of non-cubic systems, we observe a consistent reduction of the MAPE of a factor of 3-5. In particular, we observe that when the three lattice parameters display different errors, the largest errors are those that get reduced more effectively, suppressing overall the MAPE of PBE lattice constants to less than \SI{1}{\percent}.

\section{\label{sec.:results}Conclusions}
We have investigated machine learning based unit cell volume corrections for density functional theory calculations. Model agnostic supervised local explanations improve both 
PBE's and PBEsol's volume prediction of the primitive unit cell. By applying MAPLE on PBE, one achieves overall improvements on the level of PBEsol calculations, hence,
trivially reducing PBE's volume deviations from experiments by about~\SI{50}{\percent}. This is of great convenience since all large solid state databases rely on 
PBE calculations. We provide our implementation at Ref.~\cite{code}. 
Furthermore, PBEsol+MAPLE outperforms PBEsol with a roughly $1.5$ times smaller mean absolute error. The most relevant features contributing to the MAPLE models
are, indeed, given by the lattice parameters calculated with DFT, while composition-specific features are significantly less important. 
A further benefit of our approach is the implicit correction of finite temperature effects rendering time-consuming phonon calculations unnecessary. 
Since the considered experiments are mostly performed at the same (room) temperature, our trained MAPLE models are, however, not expected to generalize well to other     
temperatures. We plan to address this point in future by training on datasets with a larger temperature variation.

\backmatter

\bmhead{Acknowledgments}
This work was supported by the Volkswagen Stiftung (Momentum) through the project ``dandelion''.

\section*{Declarations}
The authors declare that there is no conflict of interest.

\newpage
\begin{appendices}
\section{Feature importances}
The importance of the features for the MAPLE models is listed in Table~\ref{tab.:feature}.
\begin{table}[h]
\begin{center}
\begin{minipage}{0.74\textwidth}
\caption{\label{tab.:feature} Most prominent feature importances for different MAPLE models. The superscript `$\ast$' indicates the 
     models using the measurement's ambient temperature
     as an additional input parameter.}
\begin{tabular}{@{}rrrrr@{}}
\toprule
& \multicolumn{4}{c}{importance [$\%$]}\\
\cmidrule(rl){2-5}
& PBE & PBEsol & PBE & PBEsol\\
feature & \multicolumn{2}{c}{+MAPLE} & \multicolumn{2}{c}{+MAPLE*}\\
\midrule
$V$ & $13.18$ & $12.61$ & $14.24$ & $14.05$\\
$b$ & $11.50$ & $10.28$ & $12.11$ & $12.83$\\
$a$ & $9.32$ & $9.91$ & $11.24$ & $10.65$\\
$c$ & $9.53$ & $9.66$ & $9.84$ & $9.82$\\
$\gamma$ & $5.61$ & $6.58$ & $5.80$ & $6.82$\\
mean melting point & $6.81$ & $6.54$ & $6.52$ & $6.67$\\
mean group & $6.07$ & $6.09$ & $4.93$ & $4.34$\\
mean block & $5.40$ & $5.36$ & $5.46$ & $5.59$\\
maximum group & $4.64$ & $4.31$ & $4.08$ & $3.92$\\
mean thermal conductivity & $3.71$ & $3.62$ & $3.01$ & $2.78$\\
mean mendeleev number & $3.37$ & $3.35$ & $3.65$ & $3.67$\\
$\beta$ & $2.63$ & $2.94$ & $2.74$ & $2.58$\\
$\alpha$ & $2.03$ & $2.39$ & $1.84$ & $2.19$\\
\bottomrule
\end{tabular}
\end{minipage}
\end{center}
\end{table}  

\section{Mapping of crystal structures}
The mapping between experimental structures stemming from ICSD and corresponding PBE structures from the materials project is listed hereafter in Table~\ref{tab.:mapping}. 

\begin{longtable}{lllr}
	\caption{Mapping between ICSD and MP entries.
\label{tab.:mapping}}\\
	\toprule
    composition & space group & MP-id & ICSD-id\\
	\midrule
\endfirsthead
	\multicolumn{4}{l}{\textsl{$\ldots$ continuation}} \\[5pt]
\endhead
	\multicolumn{4}{l}{\textsl{continuation $\ldots$}} \\
\endfoot
	\bottomrule\\
\endlastfoot
AcBr$_{3}$ & P63/m & mp-27972 & 31578\\
AcCl$_{3}$ & P63/m & mp-27971 & 31569\\
AcClO & P4/nmm & mp-30273 & 31659\\
AcH$_{2}$ & Fm$\overline{3}$m & mp-24147 & 56392\\
Ag(AuF$_{4}$)$_{2}$ & P21/c & mp-18125 & 85416\\
Ag(Mo$_{3}$S$_{4}$)$_{2}$ & R$\overline{3}$ & mp-35109 & 605576\\
Ag(Mo$_{3}$Se$_{4}$)$_{2}$ & R$\overline{3}$ & mp-1103642 & 605584\\
Ag(TeMo)$_{6}$ & C2/m & mp-29607 & 40793\\
\end{longtable}
\end{appendices}


\begin{thebibliography}{10}
\bibitem{code}
The implementation of the predictive models will be provided on github.
\end{thebibliography}
\end{document}